\def\mrec{M_{\rm rec}}
\def\mee{M_{ee}}
\def\mll{M_{\ell\ell}}
\def\eg{{\it e.g.}}

\def\etal{{\it et al.}}
\def\epem{e^+e^-}
\def\emem{e^-e^-}
\def\mz{m_Z}
\def\thetaw{\theta_W}
\def\mupmum{\mu^+\mu^-}

\def\zzh{g^{}_{ZZh}}
\def\zzhsq{g^2_{ZZh}}

\def\fbi{~{\rm fb}^{-1}}

\def\gev{\,{\rm GeV}}

\def\mz{m_Z}

\def\rts{\sqrt s}

\def\h{h}

\def\mh{m_{\h}}

\def\to{\rightarrow}
\def\lsim{\mathrel{\raise.3ex\hbox{$<$\kern-.75em\lower1ex\hbox{$\sim$}}}}
\def\gsim{\mathrel{\raise.3ex\hbox{$>$\kern-.75em\lower1ex\hbox{$\sim$}}}}

\documentstyle[12pt,equations]{article}
\def\ie{{\it i.e.}}
\def\etal{{\it et al.}}
\def\9{\phantom 0}      
\renewcommand\linebreak{\unskip\break} 
\begin{document}
\input epsf.sty
\input psfig.sty
\newlength{\captsize} \let\captsize=\small 
\newlength{\captwidth}                     

%
\font\fortssbx=cmssbx10 scaled \magstep2
\hbox to \hsize{
$\vcenter{
\hbox{\fortssbx University of California - Davis}\medskip
\hbox{\fortssbx University of Wisconsin - Madison}
}$
\hfill
$\vcenter{
\hbox{\bf UCD-98-4} 
\hbox{\bf MADPH-98-1033} 
\hbox{\bf hep-ph/9801317}
\hbox{January, 1998}
}$
}

%
\medskip
\begin{center}
\bf
Determining the Coupling of a Higgs Boson to $ZZ$ at Linear Colliders
\\
\rm
\vskip1pc
{\bf J.F. Gunion$^1$, T. Han$^{1,2}$ and R. Sobey$^1$}\\
\medskip
{\it $^1$Davis Institute for High Energy Physics}\\
{\it University of California, Davis, CA 95616}\\
{\it $^2$Department of Physics, 1150 University Avenue}\\ 
{\it University of Wisconsin, Madison, WI 53706}\\
\end{center}

\begin{abstract}
We demonstrate that, at a 500 GeV $\epem$ collider,
inclusion of  the $ZZ$-fusion process for production
of a light standard-model-like Higgs boson can substantially
increase the precision with which the $ZZh$ coupling can 
be determined (using the model-independent recoil mass
technique) as compared to employing only $Zh$ associated production.
\end{abstract}

\section{Introduction}

Once a neutral Higgs boson ($h$) is discovered, determining its coupling
to $Z$ bosons ($\zzh$) is of fundamental
importance.~\footnote{For recent reviews of Higgs boson phenomenology,
see $\eg$, Refs.~\cite{dpfreport,snowmassreport}.}
It is this coupling which most directly reflects
the role of the $\h$ in electroweak symmetry breaking.
In the minimal Standard Model (SM), where the Higgs sector consists
of a single Higgs doublet field, 
there is only one physical Higgs boson eigenstate, with
coupling $\zzh = g \mz/\cos\thetaw$, where $g$ is the $SU(2)$
coupling and $\thetaw$ is the weak mixing angle. In contrast,
in a theory with many scalar doublets and/or singlets,
the $ZZ$ couplings of the individual neutral Higgs bosons ($\h_i$)
are generally reduced in magnitude, but must obey the sum rule \cite{sumrule}
$\sum_i g_{ZZ\h_i}^2 =  [g \mz/\cos\thetaw]^2$. The sum rule becomes
still more complicated if triplet Higgs representations are present.
Precise determination of $\zzh$ for each and every observed $\h$
will therefore be crucial to knowing whether or
not we have found all the Higgs bosons that participate in electroweak
symmetry breaking and to understanding the full
structure of the Higgs sector.

In $\epem$ collisions, 
the dominant Higgs boson production diagrams involving the
$ZZ\h$ coupling are of two types:
\begin{eqnarray}
\label{BJ}
\epem &\to& Z\h  \\
\epem &\to& \epem Z^* Z^* \to \epem \h .
\label{FUSION}
\end{eqnarray}
There is (constructive) interference
of the amplitude for $\epem\to Z\h\to \epem \h$ with that for (\ref{FUSION}). 
However, it is desirable both for simplicity and in order to maximize 
experimental accuracy for the $\zzh$ determination to impose
cuts such that this interference is very small; the
$Z\h$ (\ref{BJ}) and $ZZ$-fusion (\ref{FUSION}) amplitudes
can then be considered as leading to effectively independent 
production processes.~\footnote{Nonetheless,
our calculations will always employ the full SM matrix
elements \cite{madg}, including all interfering diagrams, for 
any particular signal final state. Full matrix elements are also
employed for background processes contributing to any
particular final state. However, interference between
the signal and background is neglected; this is an excellent approximation
when considering a very narrow light Higgs boson.} The main point
of this paper is to demonstrate that, in the case
of a light Higgs boson, when the $\epem$ collider
is operated at full energy (e.g. $\rts=500\gev$), the $ZZ$-fusion production
mode will make a crucial contribution to the determination
of the $ZZh$ coupling. Indeed, by combining the $ZZ$-fusion and $Zh$
production processes we find that the error for $\zzh$ 
achieved at $\rts=500\gev$ can be competitive with that which 
is attained using $Zh$ associated
production (alone) at lower $\rts$ near the maximum in the
$Zh$ cross section. Further,
if the linear collider is run in the $\emem$ mode, the only source
of $h$ production, and only means for determining
the $ZZh$ coupling is from $ZZ$-fusion $\emem\to\emem \h$ \cite{BBCH}.

\begin{figure}[ht]
\leavevmode
\begin{center}
\epsfxsize=4.25in
\epsfysize=4.25in
\hspace{0in}\epsffile{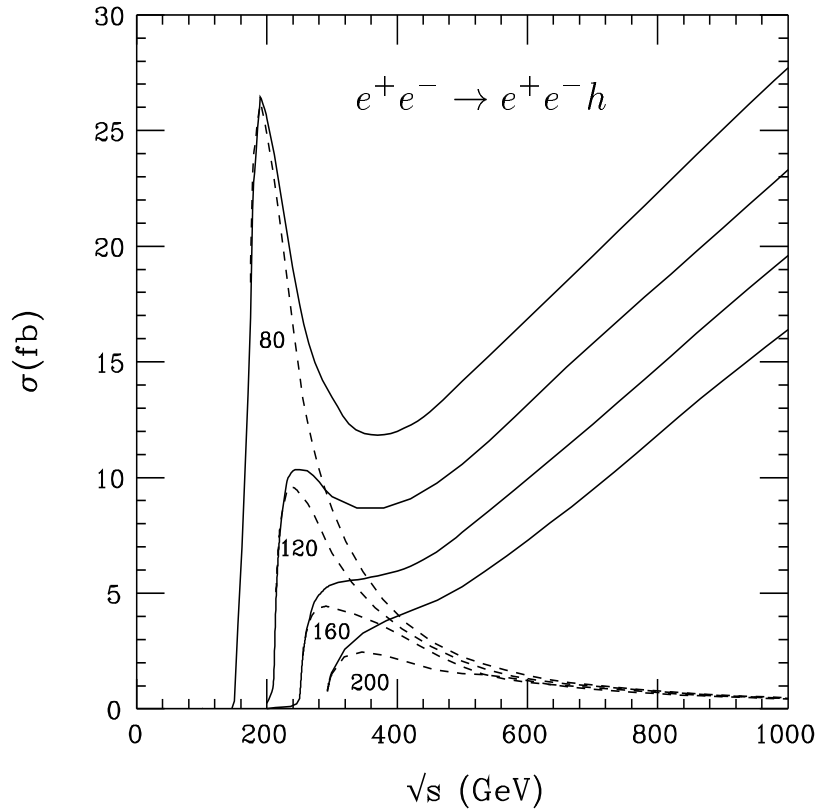}
\end{center}
\caption[]{Total cross section for 
$\protect \epem \rightarrow \epem h$ (solid) as a function of 
$\protect \rts$ for $m_h=80,120,160$ and
200 GeV. Dashed curves present the contribution only from
$\protect \epem \rightarrow Z\h$ with $\protect Z \rightarrow \epem$.}
\label{CrossSection} 
\end{figure} 

Processes~(\ref{BJ}) and (\ref{FUSION}) have quite different characteristics.
For a lighter Higgs boson and lower center of mass energy ($\rts$)
process~(\ref{BJ}) dominates (with a maximal cross section at
$\rts \sim \mz + \sqrt 2 \ \mh$), while for a heavier Higgs boson or
higher $\rts$, process~(\ref{FUSION}) becomes more important
(the cross section increasing logarithmically with $\rts$).  
This is shown in Figure~\ref{CrossSection}, where we present
the total cross section for 
$\epem \rightarrow \epem h$ (solid) as a function of 
$\rts$ for several Higgs boson masses, $m_h=80,120,160$ and
200 GeV. Dashed curves present the contribution only from
$\epem \rightarrow Z\h$ with $Z \rightarrow \epem$. Note that 
the $ZZ$ fusion cross section becomes larger than that of $Zh$
for $\rts>300$ GeV.

In both the $Z\h$ and $ZZ$-fusion 
channels, the Higgs signal can be easily detected for
$\mh\lsim (0.7-0.8)\rts$ by reconstructing
the Higgs mass peak via the $\h$ decay products.
However, in order to determine the $ZZ\h$ coupling in a {\it
model-independent} manner (\ie\ independent of the $\h$'s branching
ratio to any particular channel), it will be crucial to identify the
Higgs signal through the ``recoil mass'' variable, 
\begin{equation}
\mrec^2 = s + \mll^2 - 2\ {\sqrt s}\ (E_{\ell^+} + E_{\ell^-})\ ,
\label{RECM}
\end{equation}
where $\mll$ is the invariant mass of the final state lepton 
pair and the $E_\ell$ are the lepton energies in the c.m. frame; here,
$\ell=e,\mu$ is possible for process~(\ref{BJ})~\footnote{One could also 
consider reconstructing an $\mrec$ peak
using the $Z \to q\bar q$ (hadronic) decays in Eq.~(\ref{BJ}) in order to
increase the signal statistics. However, the energy/momentum resolution
for jets is much worse than for leptons and the backgrounds in the hadronic
decay channels are larger, implying a less sharp signal peak above
background. Thus, we will consider only the leptonic modes.}
while only $\ell=e$ is relevant for process~(\ref{FUSION}).
Due to detector resolution effects, $\mrec$ will display a peaked 
distribution of finite width, much broader than the physical width
of a light Higgs boson, centered about $\mh$. If we can measure 
the inclusive cross section associated with the $\mrec$ peak
in such a way that there is small sensitivity to the $\h$ decay, then
we can obtain a direct determination for the $ZZ\h$ coupling $\zzh$.

Since we allow the Higgs to decay to anything, our background is
composed of many processes. For $\ell=e$, for example, we must consider
all processes of the type:
\begin{eqnarray} 
\label{Background} 
\epem \to \epem X,
\end{eqnarray} 
with $X=\ell^+\ell^-, \tau^+ \tau^-, \nu\bar \nu$
and  $q \bar q$.

Many analyses of the process~(\ref{BJ}) 
in the inclusive $\mrec$ context have appeared in the literature;
see, for example, \cite{janot,kawagoe,nlcreport,snowmassreport,ecfareport}
and references therein.
However, process~(\ref{FUSION}) has received limited attention
\cite{snowmassreport,minkowski}. In particular, 
complete background computations for the inclusive
signal are given for the first time in the present paper.

The rest of the paper is organized as follows. In Sec.~2 we explore
the relative importance of the processes (\ref{BJ}) and 
(\ref{FUSION}) for determing $\zzhsq$ (assuming a SM-like $\h$), 
and comment on the $\emem\to\emem \h$ $ZZ$-fusion production mode. 
Section~3 summarizes our results and their implications for
the relative importance (for the $\zzh$ determination) of 
running at $\rts=500\gev$ vs. lowering $\rts$ to a value
near the maximum in the $Zh$ cross section.

\section[]{Determining the Coupling {\boldmath $\zzhsq$}}

In order to reconstruct the signal peak in $\mrec$ via Eq.~(\ref{RECM}), 
we must assure that the charged leptons are detected. 
Thus, we impose the following ``basic'' acceptance cuts
\begin{eqnarray}
\nonumber
|\cos\theta_\ell|<0.989, \quad p_T(\ell^\pm) > 15 \gev, \quad
 p_T(\ell^+\ell^-) > 30 \gev, \\  \mll > 40 \gev ,\ {\rm and}
\quad \mrec > 70 \gev.
\label{CUTE}
\end{eqnarray}
The polar angle cut roughly simulates the detector acceptance for
a beam hole of 150 mrad \cite{nlcreport}, and the cut on
$\mll$ is imposed in order to suppress events from photon conversion.

The sharpness of the reconstructed $\mrec$ peak is determined
by the momentum/energy resolution for the charged leptons.
The energy of an electron (but not that of a muon) can be measured in the
electromagnetic calorimeter. The momentum of either a muon or an electron
can be determined from a tracking measurement of its curvature
in the magnetic field of the detector.
We consider two possibilities for the energy resolution of the 
electromagnetic calorimeter:
\begin{itemize}
\item[I:]~NLC/EM~~  $\Delta E/E = 12\%/\sqrt E \oplus 1\%$;
\item[II:]~CMS/EM~~  $\Delta E/E = 2\%/\sqrt E \oplus 0.5\%$.
\end{itemize}
where $\oplus$ denotes the sum in quadrature and $E$ is in GeV.
Case I is that currently discussed for the NLC electromagnetic
calorimeter \cite{nlcreport}; case~II is that of the CMS lead-tungstate
crystal \cite{cms}. We also consider two possibilities for the momentum
resolution from tracking:
\begin{itemize}
\item[III:]~NLC/tracking~~  $\Delta p/p = 2 \times 10^{-4} p \oplus
1.5\times 10^{-3} /\sqrt {p}\sin^2\theta$;
\item[IV:]~SJLC/tracking~~  $\Delta p/p = 5 \times 10^{-5} p \oplus 
10^{-3}$,
\end{itemize}
with $p$ in GeV.
Case III is that specified for the typical NLC detector in \cite{nlcreport}
and case IV is that quoted for 
the ``super-JLC'' (SJLC) detector design \cite{jlci}.

It is important to note that for an electron the tracking determination
of its momentum is not statistically independent of the electromagnetic
calorimeter measurement of its energy. Thus, the two measurements
cannot be combined; one should use whichever measurement provides
the best result.  In order to provide a clean
comparison of the different possibilities, we will present results in which
we analyze all events using either the calorimeter energy measurement
or the tracking measurement; \ie\ we do not choose the best measurement on an
event-by-event basis.

It is useful to compare the fractional resolutions, $r\equiv \Delta E/E$
($E\sim p$ for $\ell=e,\mu$), for the different cases 
to one another as a function of energy/momentum.
Using $\theta=90^\circ$ in III, one finds 
that $r_{\rm III}>r_{\rm I}$ for any $E>70\gev$ (lower $E$ for $\theta$
in the forward or backward direction) and that $r_{\rm I}>r_{\rm IV}>r_{\rm
II}$ for $E>100\gev$. At $\rts=500\gev$,
lepton energies above $100\gev$ are typical
for the light Higgs masses studied here; then,
NLC/tracking will not be useful for an electron, 
whereas SJLC/tracking might be, depending upon the EM calorimeter.
Since the muon energy/momentum can only be measured by tracking,
NLC/tracking will result in the $Z\h\to \mupmum\h$ channel having a
poorer signal to background ratio than either the $Z\h\to \epem\h$ 
or the $ZZ$-fusion $\epem\h$ channel.
On the other hand, if the machine energy is lower, 
\eg\ near the peak in the $Z\h$ cross section for 
small $\mh$, electron energies are smaller, and for the majority of events
the NLC/tracking measurement of the electron energy
is competitive with the NLC/EM measurement. This will be apparent
from the figures in the next section.

To suppress the SM background more effectively and for physics
clarity, it is beneficial to divide the study into the two
natural categories: the $Zh$ associated production and $ZZ$ fusion
processes.

\subsection[]{{\boldmath $\epem \to Z\h \to \epem \h $} and 
{\boldmath $\mu^+ \mu^- \h$} }

\begin{figure}[ht]
\leavevmode
\epsfxsize=5.0in
\epsfysize=3.5in
\begin{center}
\hspace{0in}\epsffile{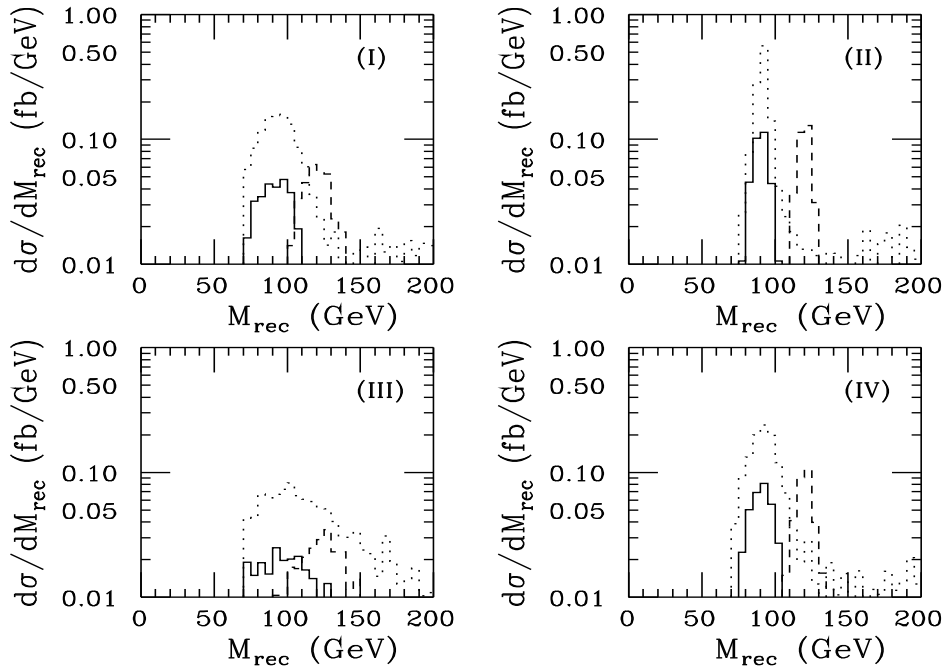}
\end{center}
\caption[]{Recoil mass distributions for $\protect\epem \to \epem X$ at
$\protect\sqrt s =500\gev$.  The solid and dashed curves
give the $Z\h$ signal for $\protect\mh=90$ and $ 120\gev$, respectively.
The dotted line is the summed SM background. Results for
the different energy/momentum resolution, cases I-IV,  are shown in the four
different panels. The cuts of Eqs.~(\protect\ref{CUTE}),
(\protect\ref{CUTMEE}) and (\protect\ref{CUTPTLL}) have been imposed.} 
\label{recmass2}
\end{figure}

We first discuss a linear collider with $\rts=500\gev$.
In order to isolate the $Z\h$ class of events we require
\begin{equation}
|\mll-\mz | < 10 \gev \ ,\quad |\cos\theta_\ell| <0.8 \ .
\label{CUTMEE}
\end{equation}
The mass cut largely eliminates the leading background from
$\epem\to W^+W^-$ and the polar angle cut helps reduce the other large
background from $\epem\to ZZ$. Since the $Z$ boson in the signal
is not only central, but also energetic with 
$E_Z^{}=(s-m_h^2+\mz^2)/2\rts \approx \rts/2$,
we can further reduce the background by imposing a cut of
\begin{equation}
p_T(e^+e^-) >80 \ \gev \ .
\label{CUTPTLL}
\end{equation}
The signal to background ratio is improved after the cuts to the
extent that the Higgs must be nearly degenerate with the $Z$
for the $\mrec$ peak to have any significant background.
Figure~\ref{recmass2} presents the recoil mass distributions, 
$d\sigma/d\mrec$, for $\epem \to \epem X$ at $\sqrt s =500\gev$.  
Results for different energy/momentum resolutions, cases I-IV,  
are shown in the four different panels.
The solid and dashed curves give the $Z\h$ signal 
for $\mh=90$ and $ 120\gev$, respectively.
The dotted line is the summed SM background.

\begin{figure}[ht]
\leavevmode
\epsfxsize=5.0in
\epsfysize=3.5in
\begin{center}
\hspace{0in}\epsffile{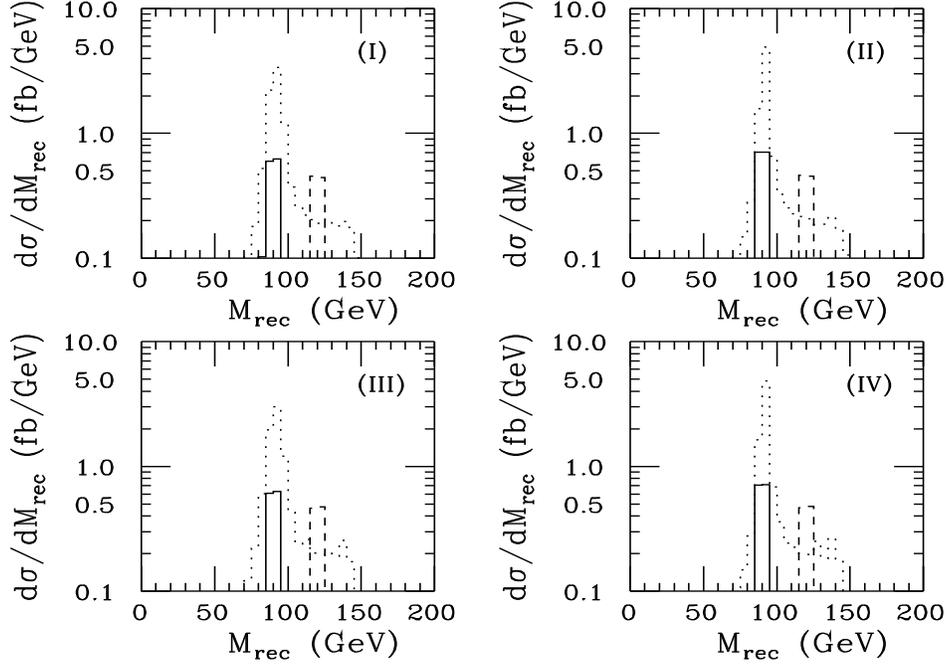}
\end{center}
\caption[]{Recoil mass distributions for $\protect\epem \to \epem X$ at
$\protect\sqrt s =250\gev$.  The solid and dashed curves
give the $Z\h$ signal for $\protect\mh=90$ and $ 120\gev$, respectively.
The dotted line is the summed SM background. Results for
the different energy/momentum resolution, cases I-IV,  are shown in the four
different panels. The cuts of Eqs.~(\protect\ref{CUTE}) and
(\protect\ref{CUTMEE}) have been imposed.} 
\label{recmass250}
\end{figure}

The $\epem \to Z\h$ cross section reaches a maximum
near $\rts \sim \mz + \sqrt 2\ \mh$. A relevant question is
how much improvement is possible by running the machine
at a lower energy nearer the maximum cross section, as would be possible
once $\mh$ is known (either from LHC or NLC data).
To illustrate, we consider $\rts=250\gev$.
Figure~\ref{recmass250} shows the recoil mass distributions, 
$d\sigma/d\mrec$, similar to those in Fig.~\ref{recmass2}
but for $\sqrt s =250\gev$. We see not only that the cross
section rate is larger, but also that the signal peak is
much sharper because of the better determination for lepton
energy/momentum at this lower $\rts$.

We estimate the relative statistical error for the cross section
measurement as
\begin{equation}
\label{ERROR}
R = \sqrt{S+B}/S \ ,
\end{equation}
where $S$ and $B$ are the numbers of 
signal and background events for a given luminosity; neglecting
systematic uncertainty from correcting for our cuts, this is also the
error on the coupling $\zzhsq$. In Table~\ref{TABZH} we compare the 
errors (as a function of $\mh$)
in the $Z\h$ mode, with $Z\to\epem$, for different
resolution choices, taking $\rts=500\gev$ and 250 GeV and assuming an
integrated luminosity of $L=200\fbi$. At $\rts=500\gev$, 
the error ranges from $\sim 15\%$ at $\mh=80\gev$
to $\sim 7\%$ at $\mh=140\gev$.
In general, the accuracy is not greatly affected by the detector. 
The exception is case III which yields poorer results than the other cases.
Results for the $\zzhsq$ error at $\rts=250\gev$ 
are $\sim 6\%$ (more or less independent of resolution choice)
for most values of $\mh$, worsening to $\sim 7\%$ at the $Z$ peak.
Thus, by running the machine at an energy near the maximum 
in the $Z\h$ cross section one should be able 
to improve the accuracy for this particular mode 
by about a factor of 2. 

\begin{table}
\begin{center}
\caption[]{
Percentage accuracy for $\protect \zzhsq$
based on $Z\h$ channel $\protect\epem\to\epem\h$ cross section
measurements with $L=200 \protect \fbi$ assuming
(a) $\protect \rts = 500 \gev$ and
(b) $\protect \rts = 250 \gev$. Results for the four different
lepton energy/momentum resolutions are shown. 
The cuts of Eqs.~(\ref{CUTE}) and (\ref{CUTMEE})
have been imposed for both $\protect \rts = 500 \gev$ 
and $\protect \rts = 250 \gev$.
For $\protect \rts = 500 \gev$ we have imposed the additional cut 
(\ref{CUTPTLL}). Note that only one kind of lepton 
is counted here.}
\label{TABZH}
\medskip
\begin{tabular}{|c|c|c|c|c|c|c|}  \hline
\multicolumn{7}{|c|}{(a)\ $\sqrt{s} = 500 \gev$}\\ \hline
{resolution} & {mass bin} & 
\multicolumn{5}{c|}{$\mh(\gev)$} \\ \cline{3-7}
$  $ & $ $ & $80$  & $90$ & $100$ & $120$  &  $140$\\ \hline
I&$(\mh \pm 10)$  & 
$15\%$ & 
$15\%$ & 
$12\%$ &  
$8.3\%$ & 
$7.2\%$  \\
II & $(\mh \pm 10)$ & 
$11\%$ &
$12\%$ &
$10\%$ &
$6.2\%$ &
$6.4\%$   \\
III & $(\mh\pm 10)$ &
$19\%$ &
$21\%$ &
$18\%$ &
$15\%$ &
$12\%$\\
IV & $(\mh \pm 10)$ &
$11\%$ &
$12\%$ &
$10\%$ &
$6.8\%$ &
$6.5\%$   \\
\hline
\multicolumn{7}{|c|}{(b)\ $\sqrt{s} = 250 \gev$}\\ \hline
I&$(\mh \pm 10)$  & 
$ 5.7\%$ & 
$ 6.7\%$ & 
$ 6.3\%$ &  
$ 4.8\%$ & 
$ 6.6\%$  \\
II & $(\mh \pm 10)$ & 
$ 5.8\%$ &
$ 6.8\%$ &
$ 6.7\%$ &
$ 4.8\%$ &
$ 6.5\%$   \\
III & $(\mh\pm 10)$ &
$5.6\%$ &
$6.7\%$ &
$6.7\%$ &
$4.7\%$ &
$6.7\%$ \\
IV & $(\mh \pm 10)$ &
$ 5.8\%$ &
$ 6.7\%$ &
$ 6.7\%$ &
$ 4.7\%$ &
$ 6.7\%$   \\
\hline 
\end{tabular}
\end{center}
\end{table}
%

\begin{table}
\begin{center}
\caption[]{
The percentage accuracy for $\protect \zzhsq$ obtained by
combining [via Eq.~(\protect\ref{comerror})] results for
the $Z\h$ cross section measurement in the
$\protect\epem\to\epem\h$ (NLC/EM) channel 
and the $\protect\epem\to\mupmum\h$ (NLC/tracking) channel,
assuming $L=200 \protect \fbi$ and $\protect \rts = 500 \gev$ 
or $\protect \rts = 250 \gev$. The cuts of 
of Eqs.~(\ref{CUTE}) and (\ref{CUTMEE}) 
have been imposed for both energies and for 
$\protect \rts = 500 \gev$ we impose
the additional cut (\ref{CUTPTLL}).
The mass bins of Table~\ref{TABZH} have been
employed.}
\label{nlcnet}
\medskip
\begin{tabular}{|c|c|c|c|c|c|}  \hline
$\rts(\gev)$ & 
\multicolumn{5}{c|}{$\mh(\gev)$} \\ \cline{2-6}
$  $ & $80$  & $90$ & $100$ & $120$  &  $140$\\ \hline
500 & $12\%$  & $12\%$ & $10\%$ & $7.3\%$ & $6.2\%$ \\
250 & $4.0\%$  & $4.7\%$ & $4.6\%$ & $3.4\%$ & $4.7\%$ \\
\hline 
\end{tabular}
\end{center}
\end{table}

We compute the net error for the $\zzhsq$ 
determination in the $Z\h$ channel by including
measurements for both the $Z\to\epem$ and the $Z\to\mupmum$
final states. The net error ($R_{\rm net}$) is given by
\begin{equation}
R_{\rm net}^{-2} = R_e^{-2} + R_{\mu}^{-2}.
\label{comerror}
\end{equation}
In obtaining the error for the $\epem$ final state, we compare results found
using the EM calorimeter to those found using tracking and
adopt the superior choice. Thus, for example, if we assume NLC/EM
and NLC/tracking, this means using case I results for the $Z\to\epem$
final state and case III results for the $Z\to\mupmum$ final state.
The results are illustrated in Table~\ref{nlcnet}.
At $\rts=250 \gev$ our results indicate that an error
for $\zzhsq$ of less than 5\% can
be expected.~\footnote{This, and other results obtained
here for the $Z\h$ mode are generally consistent with those of
Refs.~\cite{janot,kawagoe,snowmassreport,ecfareport}
for $\mh\gsim 110\gev$, when the same 
electromagnetic-calorimeter/tracking
resolution assumptions are made. For lower $\mh$, we find higher
backgrounds as compared to the estimates made in Ref.~\cite{snowmassreport},
resulting in larger errors.}

\subsection[]{ {\boldmath $\epem \to \epem \h$} 
and {\boldmath $\emem\to\emem\h$} via {\boldmath $ZZ$}-fusion}

The major advantage for the $ZZ$-fusion channel (\ref{FUSION}) 
over the $Z\h$ channel (\ref{BJ}) is that the cross section 
increases logarithmically with energy; at $\rts=500\gev$
and for $\mh=120\gev$ it is about 10 fb
as compared to the $Z\h$ cross section of about 2.5 fb. 
To remove the large $ZZ$ background, we require, in addition to 
the basic cuts of Eq.~(\ref{CUTE}),
\begin{equation}
\label{CUTZ}
\mee > 100 \gev\ .
\end{equation}
The only penalty from the $\mee$ cut is
a $20\%$ decrease in the signal rate due to elimination of the constructive
interference of the $ZZ$-fusion amplitude with the $Z\h$ amplitude.

\begin{figure}[ht]
\leavevmode
\epsfxsize=5.0in
\epsfysize=3.5in
\begin{center}
\hspace{0in}\epsffile{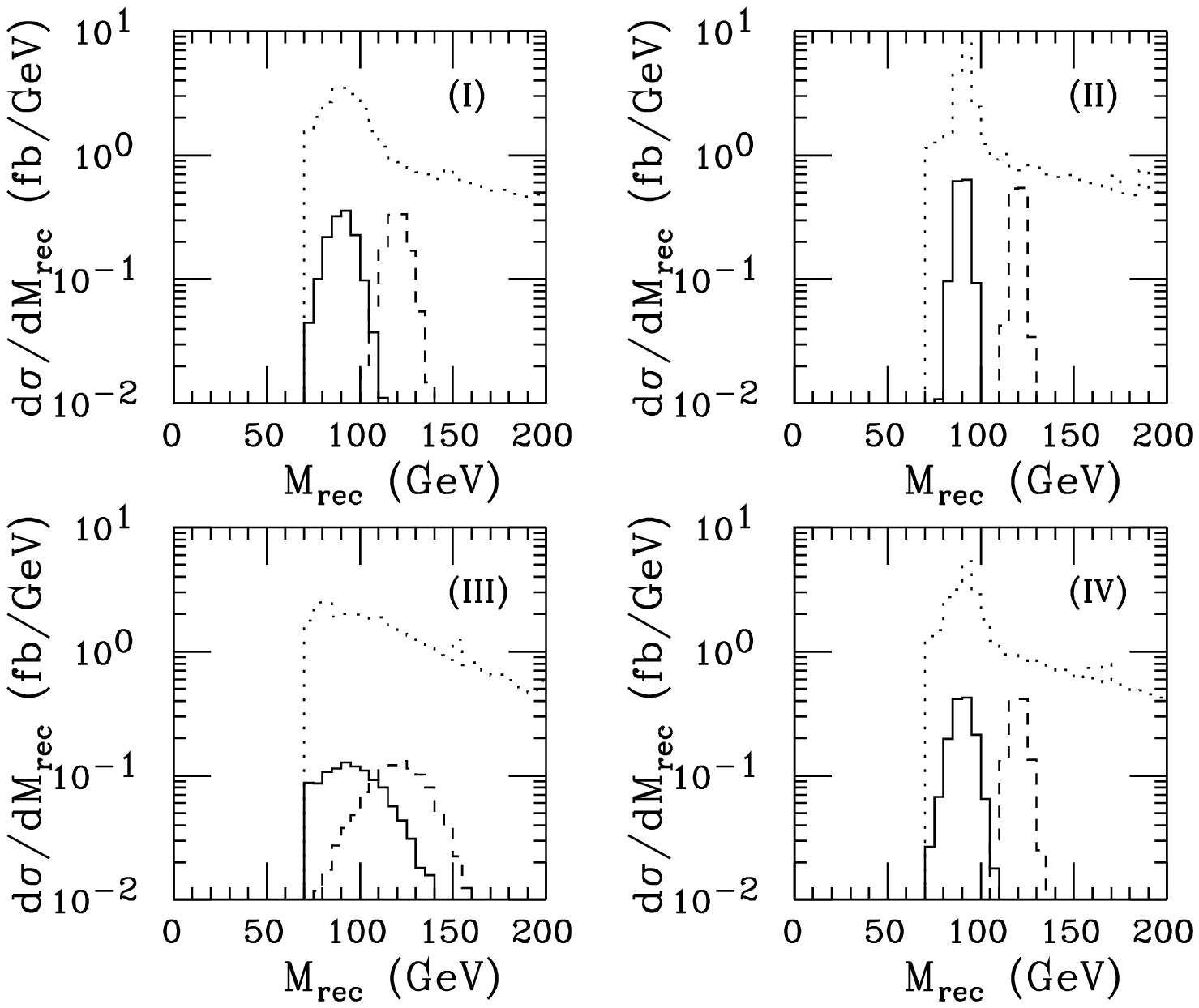}
\end{center}
\caption[]{Recoil mass distributions
for $\protect\epem \to \epem X$ at $\protect\sqrt s = 500\gev$.
Results for $\protect\mh=90\gev$ (solid), and $120\gev$ (dashed) are shown.
Results for the different energy/momentum resolution, cases I-IV, are shown 
in the four different panels.
We impose the cuts of Eqs.~(\protect\ref{CUTE}) and (\protect\ref{CUTZ}).}
\label{recmass3}
\end{figure}

Figure~\ref{recmass3} presents signal and background curves after the cuts
of Eqs.~(\ref{CUTE}) and (\ref{CUTZ}). Although some background
persists,~\footnote{The signal-to-background ratio
can be further enhanced by stronger cuts, but the signal rate
is also reduced and the error in the determination
of $\zzhsq$ is not improved.}  
the much larger signal rate makes a good measurement 
of the cross section feasible. The corresponding
results for the error of the $\zzhsq$ determination are
presented in Table~\ref{TABZFUS}. Using NLC/EM calorimetry, case I,
the error ranges from $10\%$ to $7\%$. 
As in the $Z\h$ channel, there is little difference between 
results for the resolution cases I, II and IV, whereas case III
yields substantially poorer results.

\begin{table}
\begin{center}
\caption[]{
Percentage accuracy for $\protect \zzhsq$
based on the $ZZ$-fusion channel $\protect\epem\to\epem\h$ cross section
measurements with $L=200\fbi$ at $\protect\rts=500\gev$.
Results for NLC/EM, CMS/EM and SJLC/tracking (resolution cases I, II and IV)
are shown; NLC/tracking yields much poorer results than NLC/EM.
The cuts of Eqs.~(\ref{CUTE}), and (\ref{CUTZ}) have been imposed.}
\label{TABZFUS}
\medskip
\begin{tabular}{|c|c|c|c|c|c|c|}  \hline
\multicolumn{7}{|c|}{ $\sqrt{s} = 500 \gev$}\\ \hline
{resolution} & {mass bin} & 
\multicolumn{5}{c|}{$\mh (\gev)$} \\ \cline{3-7}
$ $ & $ $ & $80$  & $90$ & $100$ & $120$  &  $140$\\ \hline
I&$(\mh \pm 10)$  & 
$9.7\%$ & 
$10\%$ & 
$9.6\%$ &  
$6.9\%$ & 
$7.2\%$  \\
II & $(\mh \pm 10)$ & 
$8.8\%$ &
$9.7\%$ &
$11\%$ &
$6.3\%$ &
$6.8\%$   \\
III & $(\mh \pm 10)$ & 
$19\%$ &
$18\%$ &
$17\%$ &
$15\%$ &
$14\%$   \\
IV & $(\mh \pm 10)$ &
$9.1\%$ &
$9.9\%$ &
$9.2\%$ &
$6.6\%$ &
$7.0\%$   \\
\hline
\end{tabular}
\end{center}
\end{table}

The ultimate accuracy for $\zzhsq$ 
at $\rts=500$ GeV
is obtained by combining the $Z\h$ and $ZZ$-fusion channel results.
In Table~\ref{ultimate}, we present the error as a function
of $\mh$, assuming that NLC/EM energy resolution is employed for
the $\epem\h$ final states of
the $Z\h$ and $ZZ$-fusion channels and that NLC/tracking
resolution is employed for the $\mupmum\h$ final state of the $Z\h$ channel.
Note that the achievable accuracy is competitive with that obtained
at $\rts=250\gev$ using the $Z\h$ mode alone 
(the $ZZ$-fusion mode being not useful at this low energy), 
especially at the larger $\mh$ values.

\begin{table}
\begin{center}
\caption[]{
Combined percentage accuracy for $\protect \zzhsq$ as obtained by including
the three measurements: $\protect\epem\to\epem\h$ for the $Z\h$
and $ZZ$-fusion (NLC/EM) plus $\protect\epem\to\mupmum\h$ 
for the $Z\h$ (NLC/tracking) at $\protect \rts=500\gev$ and $L=200 \fbi$.
Cuts as specified in Tables~\ref{TABZH} and \ref{TABZFUS} are imposed.
}
\label{ultimate}
\medskip
\begin{tabular}{|c|c|c|c|c|}  \hline
\multicolumn{5}{|c|}{$\mh(\gev)$} \\
\hline
$80$  & $90$ & $100$ & $120$  &  $140$\\ \hline
$7.5\%$ & $7.7\%$ & $6.9\%$ & $5.0\%$ & $4.7\%$ \\
\hline 
\end{tabular}
\end{center}
\end{table}

At an $\emem$ collider, Higgs production is entirely from the $ZZ$-fusion
process, $\emem\to\emem\h$.
The results presented in Table~\ref{TABZFUS} are essentially 
applicable, except that the background level is slightly smaller in
the $\emem$ case \cite{BBCH}.

\section{Summary and Conclusions}

We have investigated the precision with which the Higgs boson to $ZZ$
coupling ($\zzhsq$) can be determined by employing 
the recoil mass distribution in $\ell^+\ell^-\h$ ($\ell=e,\mu$) final states 
coming from the $Z\h$
and $ZZ$-fusion production channels at an $\epem$ collider or 
$ZZ$-fusion (alone) at an $\emem$ collider. From the 
(fully inclusive) recoil mass distribution
a direct determination of $\zzhsq$, that is independent of 
any assumptions regarding the branching ratio for the Higgs boson
to decay into any particular channel, can be made.
We considered a number of electromagnetic calorimetry and tracking resolution 
options for the detector. Results for the four options considered
are similar, except that the tracking specified for the `typical'
NLC detector yields poor results at $\rts=500\gev$.
For the calorimetry and tracking resolutions specified for this typical NLC
detector, we find the following results for the error of 
the $\zzhsq$ determination for $\mh$ in the range $80-140\gev$:
\begin{itemize}
\item
If the $\epem$ collider is run at $\rts=500\gev$, the $ZZ$-fusion
production mode yields smaller statistical error
than does the $Z\h$ production mode (even after combining
both $\epem\h$ and $\mupmum\h$ final states in the latter case).
\item 
Taking both $Z\h$ channel and $ZZ$-fusion channel measurements
at $\rts=500\gev$ and $200\fbi$ into account, the combined accuracy 
ranges from $7.5\%$ at $\mh=80\gev$ to $4.7\%$ at 
$\mh=140\gev$ (Table~\ref{ultimate}).
\item
Running at a lower energy near the $Z\h$ cross section maximum 
is possibly fruitful.  
For accumulated luminosity of $200\fbi$ at $\rts=250 \gev$, 
the error found using the (dominant) $Z\h$ production
mode is less than 5\% for $80\leq\mh\leq 140\gev$ (Table~\ref{nlcnet}).
\item
The statistical error of a measurement in the $ZZ$-fusion
$\emem\to\emem \h$ channel would be somewhat better than
that for a measurement in the $ZZ$-fusion $\epem\to\epem \h$ 
channel (Table~\ref{TABZFUS}), due to smaller backgrounds.
\end{itemize}
The most significant implication of our results is that it may not
be necessary, or even appropriate, to run at low energy in order
to obtain the best possible accuracy for the $\zzhsq$ determination.
When running at the full energy of $\rts=500\gev$,
if $\mh\gsim 120\gev$ then (for NLC/EM and NLC/tracking) the combined
$Z\h$ and $ZZ$-fusion error of Table~\ref{ultimate} is very close
to that obtained when running at $\rts=250\gev$ {\it for 
the same integrated luminosity}.  Further, it seems likely that
it will prove desirable from the point of view of other physics
to accumulate more luminosity at $\rts=500\gev$ than at $\rts=250\gev$,
and it is also possible \cite{ji} that the instantaneous luminosity
at the lower energy will be lower (assuming that the interaction
region is optimized initially for $\rts=500\gev$). In either case,
the lower energy running might not significantly improve
the $\zzhsq$ determination even if $\mh< 120\gev$.

Thus, we conclude that use of the $ZZ$-fusion mode 
(as well as the $Z\h$ mode) should provide a very
valuable increase in the accuracy that can be achieved for
the determination of the $ZZ$ coupling of a SM-like Higgs boson
at a lepton collider operating at high energy.

\section{Acknowledgements}
This work was supported in part by the DOE under contracts 
No. DE-FG03-91ER40674 and No. DE-FG02-95ER40896, 
and in part by the Davis Institute for High Energy Physics.
%
%

\clearpage
%
\end{document}